\documentclass[a4paper,11pt]{article}
\usepackage{pos}
\usepackage{float}

\usepackage{lineno}
\newcommand{\esqfig}[5]{
	\begin{figure}
		\centering
		\fbox{\includegraphics[width=#1\textwidth, height=#2\textwidth]{Figures/#3}}
		\caption{{\small #4.}}
		\label{#5}
	\end{figure}
}

\title{Monitoring the radio galaxy M87 with HAWC}
 \ShortTitle{M87 Monitoring with HAWC}

\author*[a]{T. Capistr\'an}
\author[b]{D. Avila Rojas}
\author[a]{M.M. Gonz\'alez}
\author[a]{N. Fraija}
\author[b]{R. Alfaro}
\forColl{HAWC}
\affiliation[a]{Instituto de Astronom\'ia, Universidad Nacional Aut\'onoma de M\'exico, Ciudad de Mexico, Mexico}

\affiliation[b]{Instituto de F\'isica, Universidad Nacional Aut\'onoma de M\'exico, Ciudad de Mexico, Mexico\\[0.25cm]}
 
\emailAdd{tcapistran@astro.unam.mx}
\emailAdd{daniel\_avila5@ciencias.unam.mx}
\emailAdd{magda@astro.unam.mx}
\emailAdd{nifraija@astro.unam.mx}
\emailAdd{ruben@fisica.unam.mx}

\abstract{Studies of radio galaxies at TeV energies are fascinating because their jets are misaligned concerning our sightline. Thus, it provides us with a unique opportunity to study the structure of their jets, the radiative processes, and the acceleration mechanisms involved in them. In addition, some radio galaxies have presented variability in their emission, like the giant radio galaxy M87, which has reported several activity periods. Due to its duty cycle > 95\% and instantaneous field of view of 2 sr, HAWC provides daily monitoring of variable sources visible from the Northern Hemisphere. In this work, we show the results of monitoring M87 between January 2015 and December 2018. HAWC's observations are consistent with the low activity state reported by other instruments (like H.E.S.S and MAGIC). However, after September 2017 ($\sim$~MJD 58000), the HAWC measurements of M87 show hints of higher activity.}

\FullConference{37$^{\rm{th}}$ International Cosmic Ray Conference (ICRC 2021)\\
		July 12th -- 23rd, 2021\\
		Online -- Berlin, Germany}

\begin{document}
\maketitle
%\end{document}
%=======================================================
\section{Introduction}
	\paragraph*{}The majority of bright gamma-ray sources are active galactic nuclei (AGN) which have observed with bolometric luminosities up to $10^{48}$ to $10^{49}$ ergs/s~\citep{agnhawc}. Blazars, a subclass of AGN that launches a relativistic jet aligned with our line-of-sight \citep{1979ApJ...232...34B}, are known to show large variability in the whole electromagnetic spectrum. Blazars are commonly divided in BL Lac objects and Flat Spectrum Radio Quasars (FSRQ) \citep{1996MNRAS.281..425M}.  Radio galaxies are a subclass of AGN that launch jets  oriented at large inclination angles with respect to us ($\theta > 10^\circ$) \citep{RGTorresi}. These objects give an excellent and unique opportunity to study the AGNs from another perspective owing to the ability to observe all their structure (e.g., their nucleus, jets and giant lobes). Thus, these kinds of sources help us research the physical processes of gamma-ray production and the location of the emission region.
	\paragraph*{}The radio galaxies are more difficult to detect in the gamma-ray band due to their misaligned jet, which makes the non-thermal radiation to be faint. Despite this, 30 of these sources have been detected above 100 MeV, and six of them at very-high energies (> 100 GeV)~\citep{RGTorresi}. Regardless of the low number of these sources detected in the TeV regimen, new studies of multi-wavelength observations have revealed a fast variability in the radio galaxies M87 and IC 310 (order of 4.8 min)~\citep{RGTorresi, MAGICvariability310,2016ApJ...830...81F, 2017APh....89...14F}.
	\paragraph*{}In this work, we made a study on M87 using the High-Altitude Water Cherenkov (HAWC) gamma-ray Observatory. In Section \ref{sec:TeVM87} we present a summary of the events more representative in the TeV regime of M87. In Section \ref{sec:HAWCResults} we describe the monitoring of M87 using the HAWC Observatory. Finally,  we summarise in Section \ref{sec:ConADis} the work and remark the importance of this kind of study. 
%=======================================================
\section{The Radio Galaxy M87}\label{sec:TeVM87}	
	\paragraph*{}M87 is a giant elliptical radio galaxy located in the Virgo Cluster and was the first radio galaxy detected in the TeV regimen by High-Energy-Gamma-Ray Astronomy (HEGRA) observatory in 1998~\citep{M87first}. Like other TeV radio galaxies, M87 hosts a supermassive black hole ($6 \times10^9~M_{\odot}$); it is the second nearest radio galaxy to the Earth (z= 0.0044); it is classified as Fanaroff \& Riley class I according with morphology in radio; and its relativistic jet has an angle between 15$^{\circ}$ and 25$^{\circ}$ with respect to our line of sight~\citep{Acciari2020}. 
	\paragraph*{}Radio galaxies present variability on multiple time scales, similar to Blazars~\citep{galaxies8010018, RGChallenge}. This is the reason gamma-ray observatories started to monitor these kinds of sources. The results on M87 campaigns reports three flaring states during the years 2005, 2008 and 2010~\citep{RGATTeV}. After 2010, it has been in a quiescent or low-emission state.  It was reported by MAGIC observatory and also observed from gamma-ray to radio band~\citep{Acciari2020}. The flux normalization at 1 TeV of this source in a quiescent state is between $[2-7] \times 10 ^{-13}$~cm$^{-2}$~s$^{-1}$~TeV$^{-1}$ reported by MAGIC, VERITAS and H.E.S.S. and during the flaring state is around $[1-5] \times 10^{-12}{\rm cm^{-2} s^{-1}TeV^{-1}}$. In both cases, the spectral model used is a simple power-law function with an index of $\sim$2.3 with an exception for the observation of 2004 by H.E.S.S. and 2012 by VERITAS, where they used a spectral of $\sim$2.6.
	\paragraph*{}HAWC is an observatory that operates almost all day (duty cycle > 95\%), so it makes a perfect instrument to monitor the source every day. Another critical point of this experiment is the wide field of view that can detect sources with declination from -26~$^\circ$ (a small part of the South sky) to 64$^\circ$ (North sky), which M87 is inside of it. Another more is the detection of gamma-ray at TeV~\citep{HAWC2017}.
% 	The disadvantage of Cherenkov Telescopes (like MAGIC or H.E.S.S. ) is that they only operate under special conditions, which means their duty cycle is low. 
%=======================================================
\section{HAWC results}\label{sec:HAWCResults}
	\paragraph*{}The analysis performed to obtain the light curve of M87 is the same as Abeysekara et. at~\citep{monitorhawc} used for reporting the daily monitoring of two blazars and the Crab Nebula of the first 17 months of HAWC. This analysis is computed with the Likelihood fitting Framework (LIFF) package\citep{LIFFcita}. This package estimates the flux and significance using the maximum likelihood method to fit a physical spectral model convolved with the HAWC detector response to the data.
	\paragraph*{}The daily light curve is obtained using HAWC data from January 1st 2015 to December 8th 2018. A simple power-law spectrum model is used, with a pivot energy of 1 TeV and an index of 2.3. The Franceschini model for EBL attenuation is also used. Two cuts were applied on the light curve: first, a minimum exposure of 4.2 hours, and second, only plot days with a significance greater than 2~$\sigma$. Figure \ref{Fig:FluxDaily} shows the daily light curve using the previous quality cut. During this period,  the fluxes of M87 have a value between [0.3-1.0]~$\times 10^{-11}$~ph~cm$^{-2}$~s$^{-1}$. 
	\esqfig{0.9}{0.5}{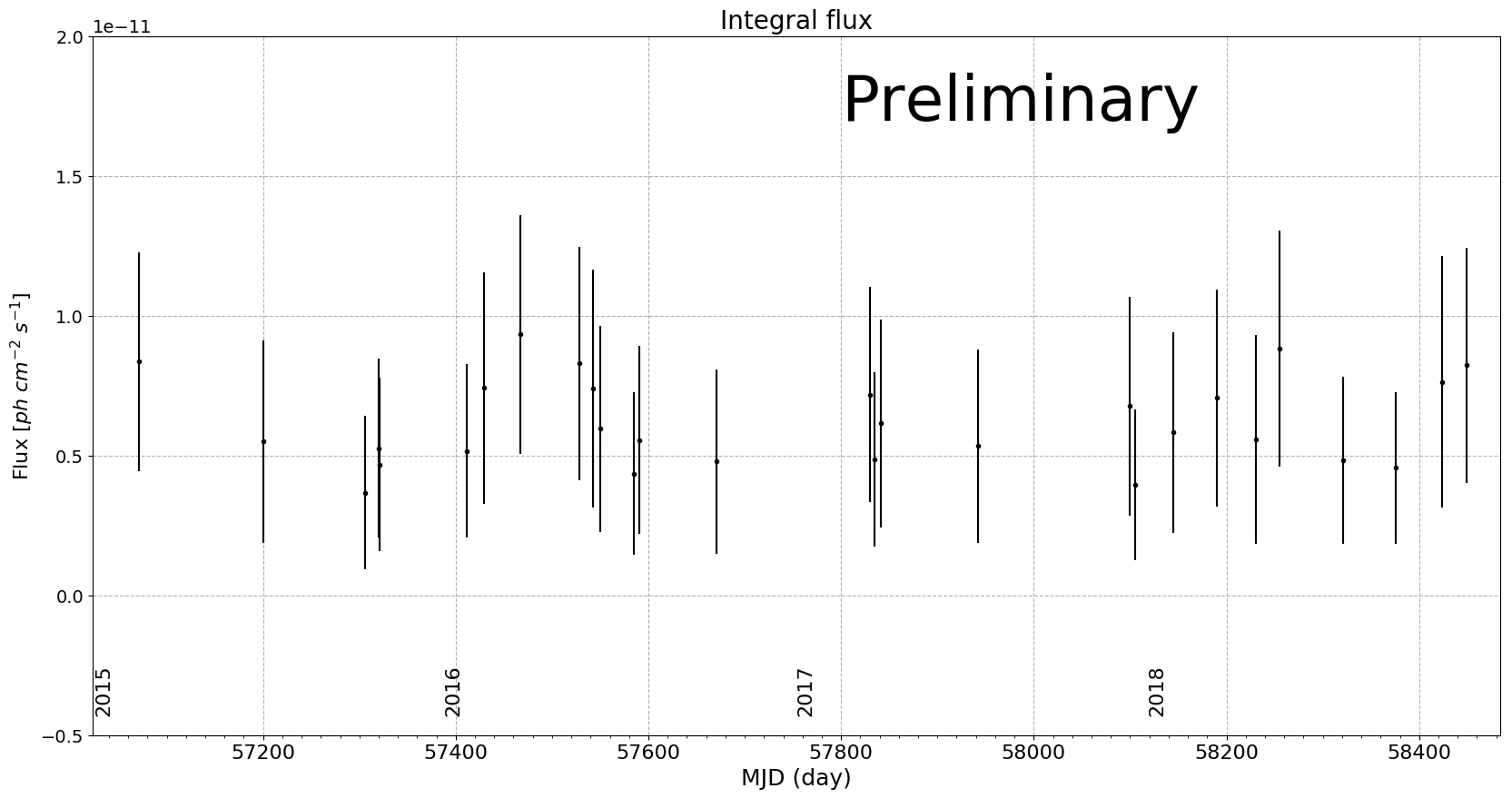}{Daily light curve of M87 from January 1st 2015 (MJD~57023) to December 8th 2018 (MJD~58460). Here only are shown all days with a significance greater than 2}{Fig:FluxDaily}
	\paragraph*{}Figure \ref{Fig:FluxMonth} show the monthly light curve. This curve uses the same period of HAWC data and the same spectral model than the daily light curve. The main difference in this plot is the calculation of the upper limits at the confidence level of  95\%. The integral flux point and the upper limit values are consistent with a non-active state. These results on M87 are consistent with the quiescent state observed in this radio galaxy. 
	\esqfig{0.9}{0.5}{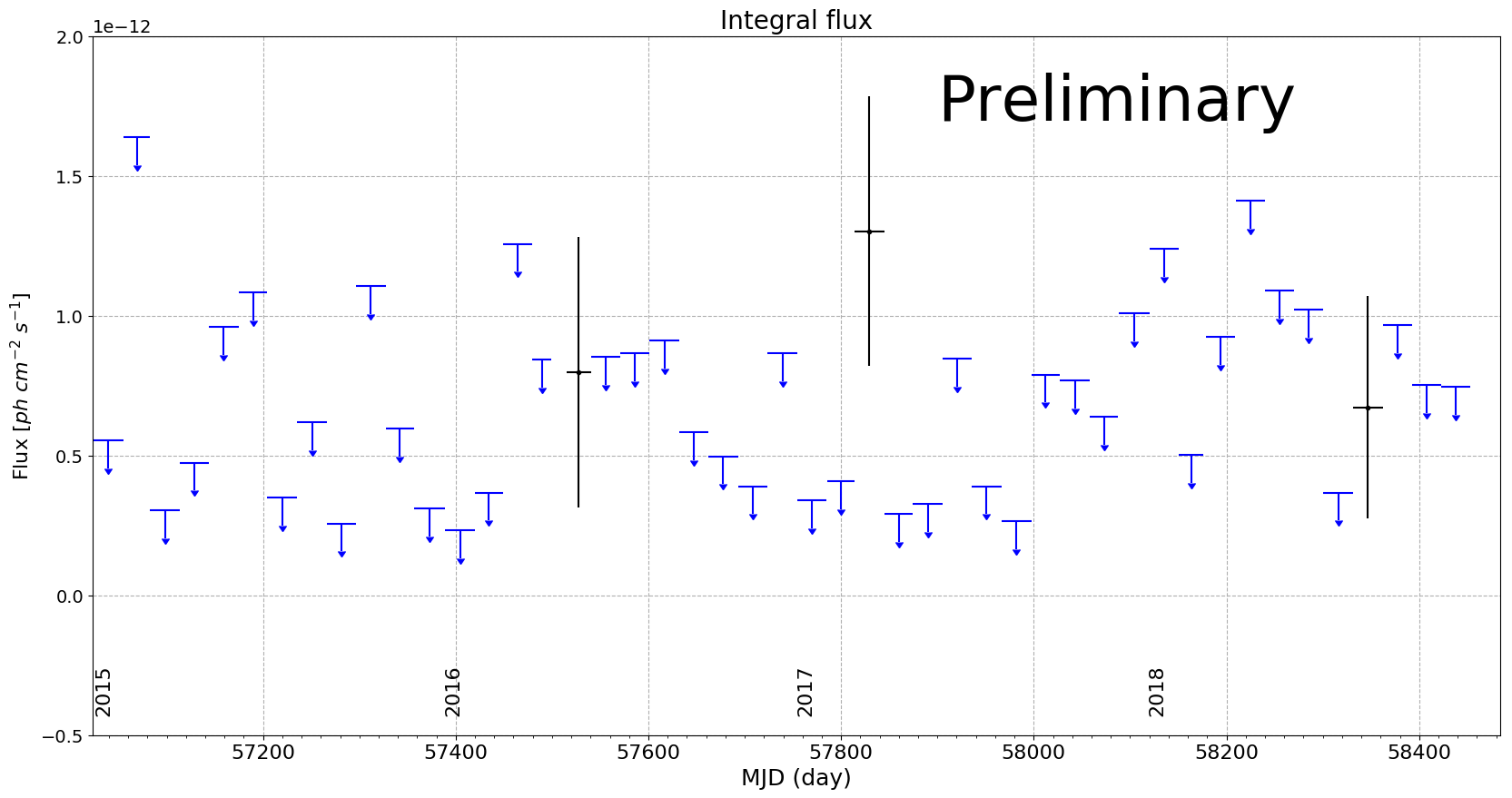}{Monthly light curve of M87 from January 1st 2015 (MJD~57023) to December 8th 2018 (MJD~58560). The black marker are the months that has a significance greater than 2~$\sigma$, the rest, blue ones, are the upper limit at 95\%}{Fig:FluxMonth} 
	\paragraph*{}We then looked at integrating the data around M87 over time. A cumulative significance was calculated adding a month by a month until the four years were covered and it is shown in Figure \ref{Fig:SigAcum}. We can see that after September 2017 (MJD 58000) the significance starts to rise steadily. Using the full livetime of the analysis, the final significance obtained is 1.8~$\sigma$, , while restricting the livetime by using data above MJD 58000, the significance increases up to 2.6~$\sigma$. 
	\esqfig{0.9}{0.5}{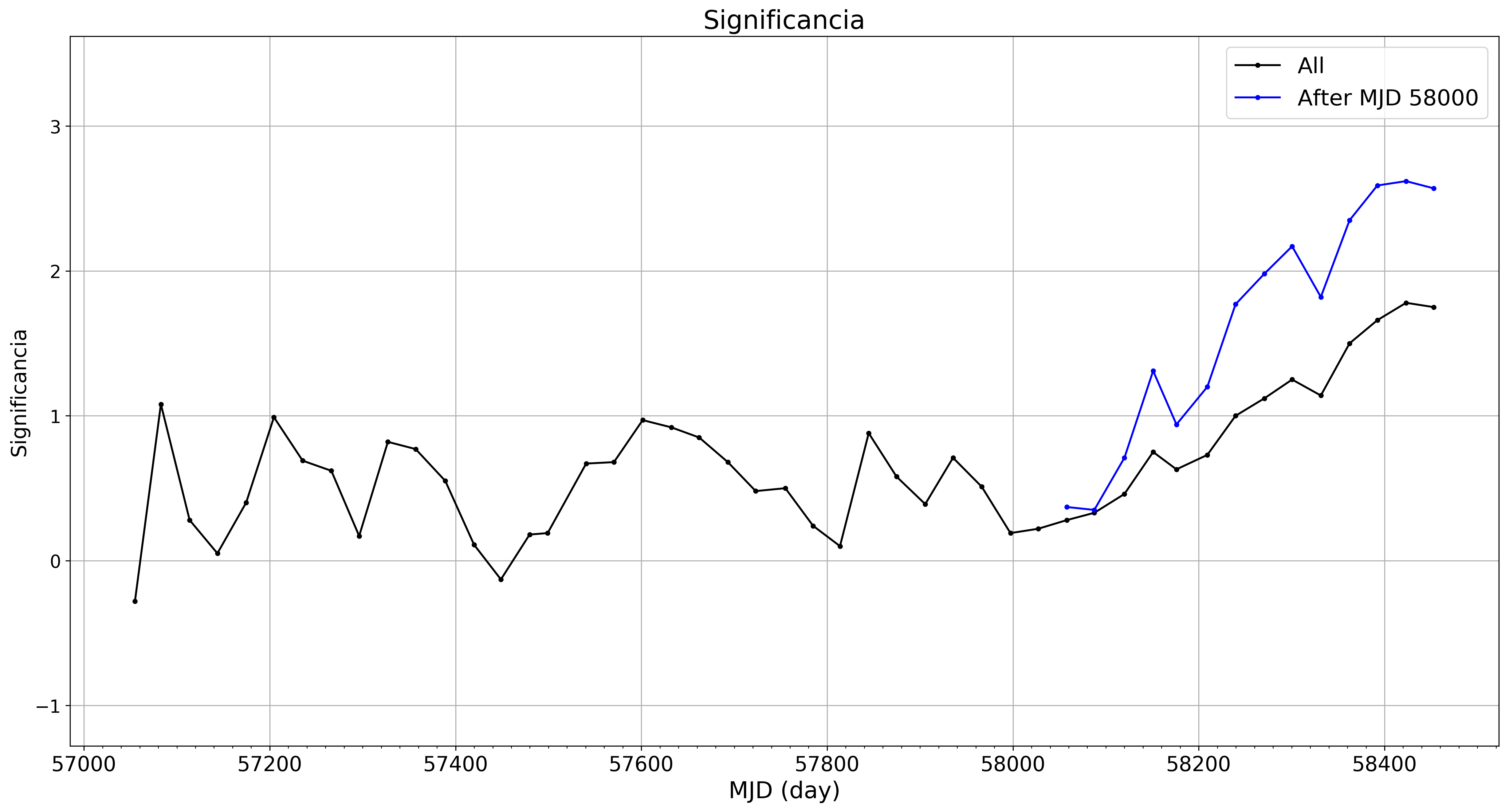}{Cumulative significance is shown by adding month-by-month data from January 2015 (MJD~57023) to December 2018 (MJD~58560). After September 2017 (MJD 58000), the significance starts to increase}{Fig:SigAcum}
%=======================================================
\section{Conclusion}\label{sec:ConADis}
	\paragraph*{}M87 has been monitored using four years of HAWC data with two different time scales. During this period of observation, M87 was found to be in a nonactive state. However, by restricting the live-time using data after September 2017, we observe a hint of evidence of TeV emission from M87. It could mean that this radiogalaxy is becoming active again. HAWC will keep monitoring the evolution of this source.
	\paragraph*{}In this work, the significance obtained in the radio galaxy M87 is around 2~$\sigma$ using around four years of data. New improvements on the analysis are under way and some preliminary results show that this radio galaxy is now significant. We will present this in the future. 
\acknowledgments{
We acknowledge the support from: the US National Science Foundation (NSF); the US Department of Energy Office of High-Energy Physics; the Laboratory Directed Research and Development (LDRD) program of Los Alamos National Laboratory; Consejo Nacional de Ciencia y Tecnolog\'ia (CONACyT), M\'exico, grants 271051, 232656, 260378, 179588, 254964, 258865, 243290, 132197, A1-S-46288, A1-S-22784, c\'atedras 873, 1563, 341, 323, Red HAWC, M\'exico; DGAPA-UNAM grants IG101320, IN111716-3, IN111419, IA102019, IN110621, IN110521; VIEP-BUAP; PIFI 2012, 2013, PROFOCIE 2014, 2015; the University of Wisconsin Alumni Research Foundation; the Institute of Geophysics, Planetary Physics, and Signatures at Los Alamos National Laboratory; Polish Science Centre grant, DEC-2017/27/B/ST9/02272; Coordinaci\'on de la Investigaci\'on Cient\'ifica de la Universidad Michoacana; Royal Society - Newton Advanced Fellowship 180385; Generalitat Valenciana, grant CIDEGENT/2018/034; Chulalongkorn University’s CUniverse (CUAASC) grant; Coordinaci\'on General Acad\'emica e Innovaci\'on (CGAI-UdeG), PRODEP-SEP UDG-CA-499; Institute of Cosmic Ray Research (ICRR), University of Tokyo, H.F. acknowledges support by NASA under award number 80GSFC21M0002. We also acknowledge the significant contributions over many years of Stefan Westerhoff, Gaurang Yodh and Arnulfo Zepeda Dominguez, all deceased members of the HAWC collaboration. Thanks to Scott Delay, Luciano D\'iaz and Eduardo Murrieta for technical support.
}
\bibliographystyle{JHEP} 
\bibliography{biblio}

%% Full authors list (ONLY FOR COLLABORATIONS)
\clearpage
\section*{Full Authors List: \Coll\ Collaboration}
%\noindent \textbf{Note comment afterwards:} Collaborations have the possibility to provide an authors list in xml format which will be used while generating the DOI entries making the full authors list searchable in databases like Inspire HEP. For instructions please go to icrc2021.desy.de/proceedings or contact us under icrc2021proc@desy.de.\\
\scriptsize
\noindent
A.U. Abeysekara$^{48}$,
A. Albert$^{21}$,
R. Alfaro$^{14}$,
C. Alvarez$^{41}$,
J.D. Álvarez$^{40}$,
J.R. Angeles Camacho$^{14}$,
J.C. Arteaga-Velázquez$^{40}$,
K. P. Arunbabu$^{17}$,
D. Avila Rojas$^{14}$,
H.A. Ayala Solares$^{28}$
% ...

\noindent
% ...
$^{14}$Instituto de Física, Universidad Nacional Autónoma de México, Ciudad de México, México,
% ...
$^{17}$Instituto de Geofísica, Universidad Nacional Autónoma de México, Ciudad de México, México,
% ...
$^{21}$Physics Division, Los Alamos National Laboratory, Los Alamos, NM, USA,
% ...
$^{28}$Department of Physics, Pennsylvania State University, University Park, PA, USA,
% ...
$^{40}$Instituto de Física y Matemáticas, Universidad Michoacana de San Nicolás de Hidalgo, Morelia, Michoacán, México,
$^{41}$FCFM-MCTP, Universidad Autónoma de Chiapas, Tuxtla Gutiérrez, Chiapas, México,
% ...
$^{48}$Department of Physics and Astronomy, University of Utah, Salt Lake City, UT, USA,
% ...

\end{document}